\def\cm2{cm$^{-2}$}

\def\c2{C~{\sc ii}}
\def\c4{C~{\sc iv}}
\def\fe2{Fe~{\sc ii}}
\def\fe3{Fe~{\sc iii}}
\def\mg1{Mg~{\sc i}}
\def\mg2{Mg~{\sc ii}}
\def\si2{Si~{\sc ii}}
\def\si4{Si~{\sc iv}}
\def\al2{Al~{\sc ii}}
\def\al3{Al~{\sc iii}}
\def\o1{O~{\sc i}}
\def\n1{N~{\sc i}}
\def\h1{H~{\sc i}}

\def\approxlt{\mathrel{\spose{\lower 3pt\hbox{$\sim$}}
        \raise 2.0pt\hbox{$<$}}}
\def\approxgt{\mathrel{\spose{\lower 3pt\hbox{$\sim$}}
        \raise 2.0pt\hbox{$>$}}}

\documentclass[article]{has40}
\usepackage{graphicx}
\usepackage{amssymb}
\tabletypesize{\normalsize}  

\def\plotone#1{\centering \leavevmode
\includegraphics[width=.95\columnwidth]{#1}}

\def\plotone#1{\centering \leavevmode
\includegraphics[width=.95\columnwidth]{#1}}

\shortauthors{Smith}
\shorttitle{Period Changes}

\begin{document}
\large    
\pagenumbering{arabic}
\setcounter{page}{1}

\title{PERIOD CHANGES OF MIRA VARIABLES, RR LYRAE STARS, AND TYPE II CEPHEIDS}

%
%
\author{{\noindent Horace Smith{$^{\rm 1}$}\\
\\
{\it (1) Dept. of Physics and Astronomy, Michigan State University, \\East Lansing, MI, USA 48824} 
}
}

%
%
\email{(1) smith@pa.msu.edu}


\begin{abstract}
Mira variables, RR Lyrae variables, and type II Cepheids all represent evolved states of low-mass stars, and long
term observations have revealed that changes in pulsation period occur for each of these classes of variable.
Most Mira variables show small or no period changes, but a few show large period changes that can plausibly be associated
with thermal pulses on the asymptotic red giant branch.  Individual RR Lyrae stars show period changes that do not
accord with the predictions of stellar evolution theory. This may be especially true for RR Lyrae stars that exhibit the Blazhko effect.  However, the average period changes of all of the RR Lyrae
variables within a globular cluster prove a better but still
imperfect match for the predictions of evolutionary theory.
The observed period changes of short period type II Cepheids (BL Her stars) as well as those of long period type
II Cepheids (W Vir stars) are in broad agreement with the rates of period changes expected from their evolutionary
motions through the instability strip.
\end{abstract}

\section{Historical Introduction}

That certain variable stars change in brightness in a periodic way was well established by the beginning of the
19th century. Mira, the archytype of its class of variable stars, was discovered in 1596, and by the mid 17th century
it was known that its brightness varied with a period of about 333 days. Goodricke and Pigott discovered the first
Cepheid variables, $\eta$ Aquilae and $\delta$ Cephei, in the late eighteenth century (French 2012).  Goodricke and
Pigott are
also responsible for the discovery of the periodic nature of the dimmings of the eclipsing binary Algol
(French 2012). Once the existence of 
periodicities in stellar variability was established, it was not long before the question arose as to whether
those periods were fixed or mutable.

It was soon obvious to observers of Mira itself and of other bright red variables that their
light curves did not repeat perfectly from cycle to cycle (Figure 1).  To account for at least some of these differences,
some early researchers attempted to
apply additional periodic terms to the ephemeris, implying that the periods of such stars were themselves variable in a
periodic way.
A number of ephemerides of this type can be found, for example, in Chandler's (1896) Third Catalogue of Variable Stars.
Disappointment followed when continued observations showed that the stars did not follow these ephmerides.  The apparent
period changes did not repeat in a periodic fashion.

\begin{figure*}
\centering
\plotone{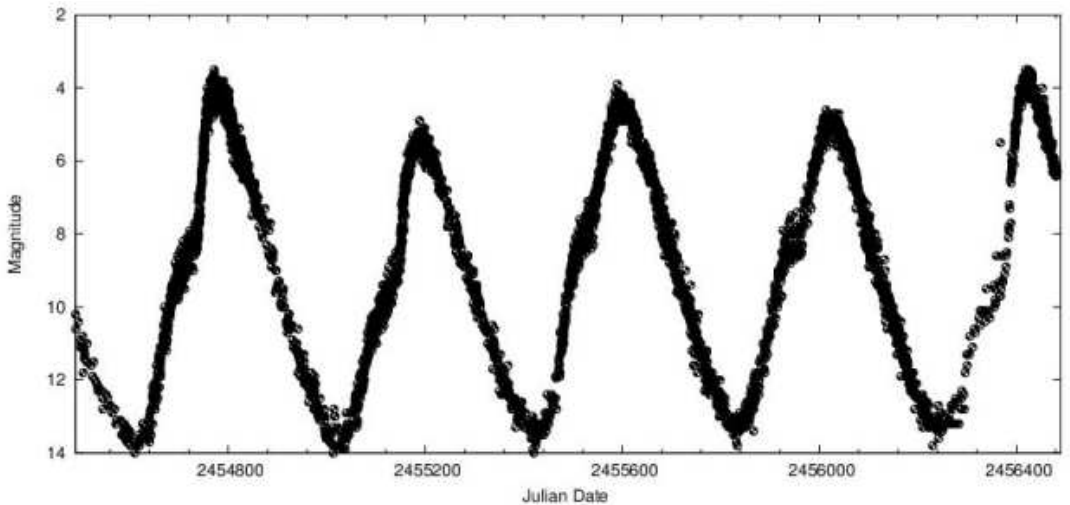}
\vskip0pt
\caption{Visual observations of the 408 day period Mira variable $\chi$ Cygni, based
upon several years of observations from the AAVSO International Database
(Henden 2013). }
\end{figure*}

In this paper we review what is known of the period changes of three types of variable star: Mira variables, 
RR Lyrae stars, and type II
Cepheids.  All three of these variables represent evolved stages in the lives of relatively low mass stars. They have,
however, very different ranges of pulsation period.  Mira variables have periods longer than 100 days, typically around
300 days.  RR Lyrae stars pulsate with principal periods usually in the 0.2 to 0.9 day range.  Type II Cepheids have
periods of 1 day to about 25 days. In
considering the period changes of these variables, it will be useful to compare the
length of time during which the variables have been observed to two timescales intrinsic to the behavior
of the star: the duration of the primary pulsation cycle and the time interval during which significant changes in period
would be expected according to our current understanding of stellar evolution.  In fact, once theories of stellar
evolution began to be developed, it was not long before astronomers sought to employ period changes to confirm
or refute the predicted timescales of stellar evolution.

Shapley (1914) argued that the light variations of Cepheids were due to pulsation, rather than being some consequence
of Cepheids being binary stars, as had often been previously supposed. The realization that Cepheids and related
variable stars were pulsating meant that they should obey the basic pulsation equation: $P\sqrt{\rho} = constant$.
Eddington (1918) noted that, applying that equation, one could use the rate of period change in a Cepheid to 
determine its rate of physical evolution. The following year, Eddington (1919) used observations of $\delta$ Cephei
extending over 126 years to show that its observed rate of period change was much smaller than would be expected
if it derived its energy solely from gravity via the Kelvin-Helmholtz contraction mechanism.

\section{Mira Variables}

Mira itself was discovered in 1596, and a small number of the brighter Mira variables were discovered in the following
two centuries.  The observational histories of even these bright Mira variables often have significant
gaps until the latter part of the nineteenth century (see, for example, the discussion of the period history
of $\chi$ Cygni by Sterken et al. 1999).  Until recent years, observations of the brighter Mira
variables were obtained mainly by visual methods.  Because of the long periods of Mira variables, we often have
observational records extending over fewer than 150 cycles, even for those Mira variables that have been
well observed since 1900.

Early studies of period changes of Mira variables (e.g. Eddington \&
Plakidis 1929; Sterne \& Campbell 1937) attempted to distinguish significant period changes from random cycle-to-cycle
period fluctuations using as a tool the $O-C$ diagrams for the time of maximum light.  This approach has been adopted in
some more recent investigations of Mira period changes (Percy et al. 1999a, b), but other approaches are now
also possible. Templeton et al. (2005) used
a form of wavelet analysis developed by Foster (1996). The Templeton et al. (2005)
study remains the most comprehensive study of the period changes of Mira stars, and we will use it as the basis
for our consideration of their period changes.

Templeton et al. (2005) investigated the period changes of 547 Mira variables having extensive observations in the
AAVSO International Database.  The average period of these Mira stars is 307 days.  57 stars, 21 stars, and 
8 stars had values of $ d ln P/dt$ that were significant at the 2 $\sigma$, 3 $\sigma$, and 6 $\sigma$ levels,
respectively. The period changes of the eight Mira variables with the most significant period changes have also been identified in other period change studies.  Figure 2 (courtesy of M. Templeton, private communication 2013)
shows the derived values of period versus time for T UMi, one of the rare Mira variables with a dramatically
changing period. The details of the period change behavior of the 8 stars with the most significant period changes
are not, however, identical. Both period increases and period decreases occur.

 It has been argued (Wood \& Cahn 1977; Wood \& Zarro 1981; Templeton et al. 2005 and references therein) that those Mira variables undergoing large period
changes are asymptotic red giant branch stars undergoing thermal pulses. The 1.6 percent of the Miras in their
sample with large period changes is consistent with the 1 or 2 percent expected to be experiencing a shell flash event. It is not clear, however, that all large period changes in Mira variables can be attributed to thermal pulses. Zijlstra et al. (2004), noting that the large increase in the period of BH Cru (420 days to 540 days) was associated with changes in its spectrum, proposed that BH Cru underwent a decrease in effective temperature not necessarily caused by a thermal pulse.

\begin{figure*}
\centering
\plotone{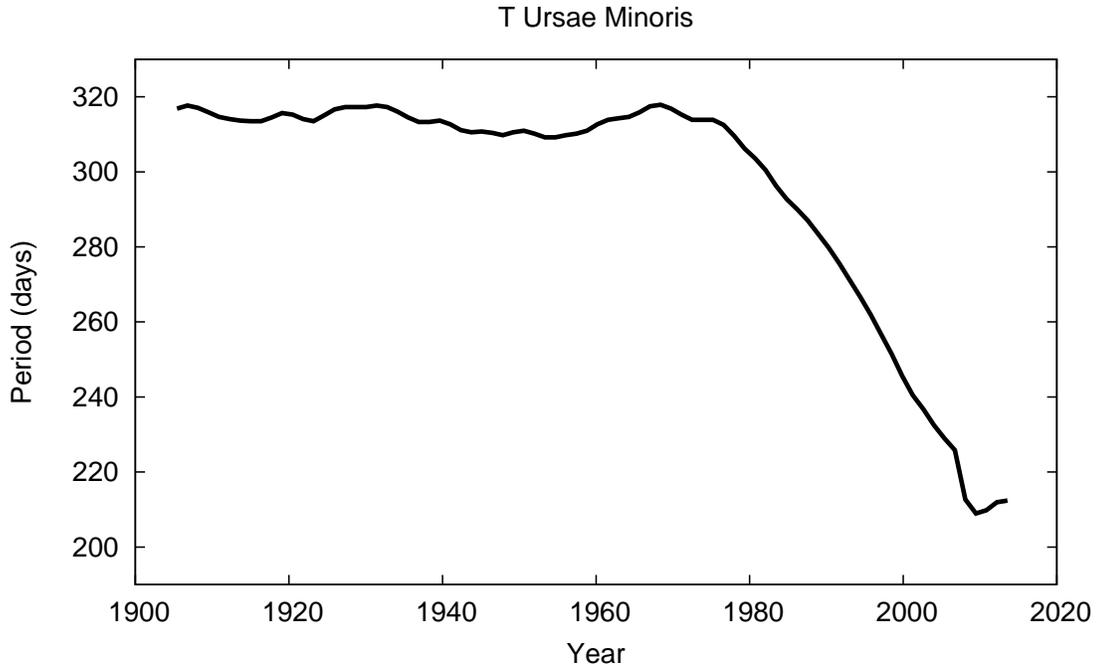}
\vskip0pt
\caption{This figure, an update of figure 8 in Templeton et al. (2005), shows the drastic decrease in the pulsation period of the Mira
variable T UMi that may be associated with a thermal pulse. }
\end{figure*}

Templeton et al. (2005) did not detect significant period changes for most of the Mira stars in their sample, a result consistent with the expectation of low rates of period change for Mira stars not undergoing thermal pulses.  Nonetheless, smaller but still significant period changes were identified in stars other than the eight thermal pulse candidates. Several Mira variables had non-monotonic period changes, appearing to oscillate around a mean period
or around a period trend.  Zijlstra \& Bedding (2002) called such stars ``meandering Miras''. Templeton et al.
noted that, while the physical mechanism for such period meandering was unclear, the timescales of these period changes
were consistent with the Kelvin-Helmholz cooling time of the envelope. They suggested that period changes of this sort might
be caused by thermal relaxation oscillations in the stellar envelope, perhaps in response to the global changes caused
by a thermal pulse.  

Uttenthaler er al. (2011) searched
for the short lived radioactive element technetium in the spectra of 12 Mira variables, noting that its presence
would be indicative of the mixing expected during a thermal pulse. Technetium was detected in five of the stars,
but they found no strong correlation between the presence or absence of technetium and the type of period change
observed within the Mira variable.

\section{RR Lyrae Stars}

\subsection{The Period Change ``Noise''}
The first field and cluster RR Lyrae stars were identified toward the end of the 19th century (Smith 1995).  Studies of RR Lyrae
variables in clusters, led by the studies of Solon Bailey and his collaborators (Smith 2000), were intially most important
in defining the characteristics of RR Lyrae stars, which were often termed ``cluster variables'' during the first half
of the twentieth century.  The gradual accumulation of observations spanning several decades 
revealed
that some RR Lyrae variables undergo significant changes in period
(e.g. Leavitt \&
Luyten 1924; Florja 1931; Martin 1938; Prager 1939). The timespan now available for many RR Lyrae
variables, about 100 years, is comparable to that of many Mira variables.  However, the much shorter pulsation period
for RR Lyrae variables, about half a day, means that during that interval some 70,000 cycles have elapsed, many more than
the 150 or so experienced by Mira variables.

RR Lyrae stars are believed to be horizontal branch (HB) stars burning helium in their cores, and one would expect this to be
reflected in their period changes.  According to stellar evolution theory, a lower mass star spends about $10^{8}$ years
on the horizontal branch, though not all of that time need be spent within the instability strip. Calculations of the
period change rates of RR Lyrae stars that would be expected just from the rate of nuclear burning
(see, for example, Sweigart \& Renzini 1979,
Koopmann et al. 1994, and Kunder et al. 2011) make two general, testable
predictions: (1) until the very end of the HB lifetime, the rate of period change should be small, less than about 0.1 day
per million years; and (2) few RR Lyrae stars should be observed to significantly alter the rate and sign of their
period change during an interval as short as a century. Enlarging upon (1), large
evolutionary period increases would be expected only for the few RR Lyrae stars evolving to the red toward the very
end of their HB lifetime.  Large evolutonary period decreases should occur even more rarely. Only
a very small number of RR Lyrae that have not yet reached the zero-age horizontal branch (ZAHB) or those
undergoing brief structural instabilities toward the end of the HB lifetime might show
large period declines (Silva Aguirre et al. 2008; Sweigart \& Renzini 1979; Koopman et al. 1994). Observations reveal a different picture.

While some field RR Lyrae stars have been constant in period for decades, other have changed signficantly in period (e.g. Tsesevich 1966;
LeBorgne et al. 2007; Percy \& Tan 2013). Figure 3 shows the $O-C$ diagram of the field RRab star XZ Cyg using data from the GEOS database (LeBorgne et al. 2007).  From this
plot we can see that the period of XZ Cyg has changed, with the period undergoing both increases and decreases
over the past century.  These changes are far larger than the observational uncertainty of the period of the 
variable. Baldwin \& Samolyk (2003) used 38 years of AAVSO data for XZ Cyg to identify seven distinct time
intervals, in each of which the star was characterized by a significantly different period. A perusal of the GEOS $O-C$
diagrams confirms that XZ Cyg is far from unique in having observed period changes that depart from those expected were
the period changes determined solely by changes in the structure of the star wrought by nuclear burning.  The observed period changes are
often too large, too abrupt, and too variable in direction to match theoretical predictions. The same is true for RR Lyrae stars within
globular clusters, as will be discussed in more detail below. It is worth noting that, even in those RR Lyrae stars
that do have large observed period changes, the changes are still relatively small compared to the periods themselves.  
Aside from a few RR Lyrae
stars that may alternate between pulsation modes (e.g. V79 in M3; Clement \& Goranskij 1999), the observed period changes 
are in all or almost all cases smaller than a few parts in $10^4$.

\begin{figure*}
\centering
\plotone{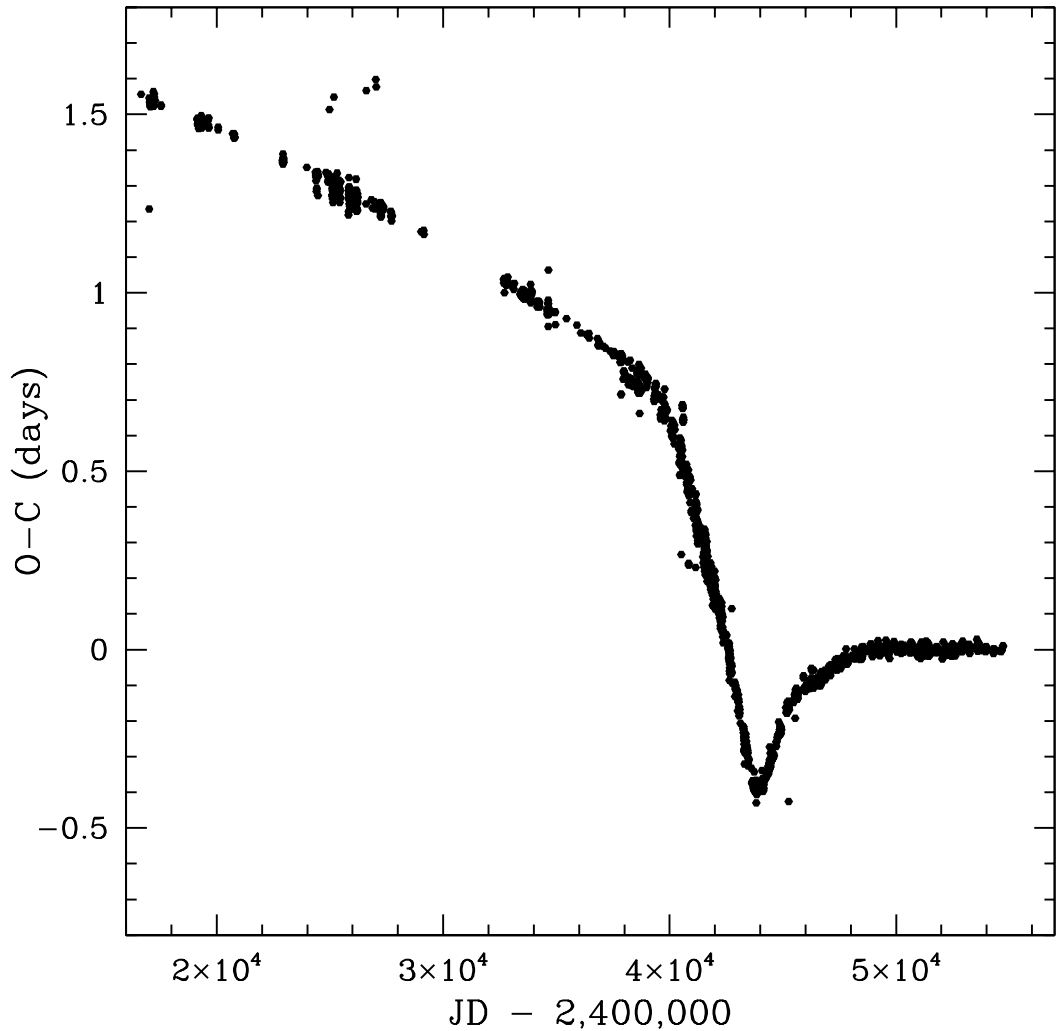}
\vskip0pt
\caption{The O-C diagram of the RRab star XZ Cygni, calculated assuming a constant period and employing
data in the GEOS database.  XZ Cyg, a star which exhibits the Blazhko
effect, has clearly undergone several changes in period of different magnitude
and sign. }
\end{figure*}

The long term period change rates of RR Lyrae stars are often characterized by the quantities $\beta = dP/dt$ per million
years or $\alpha = (1/P)(dP/ dt)$ per million years.  Because the $O-C$ diagram of a variable star with a period slowly changing
at a constant rate is well approximated by a parabola, values of $\beta$ and $\alpha$ are often determined by fitting a parabola
to the $O-C$ diagram. Figure 3 makes us aware, however, that parabolas might in fact poorly represent the period changes
of many RR Lyrae stars.

If stellar evolution theory for HB stars is correct, RR Lyrae stars need to have some source of period change noise
superposed on their evolutionary period change to account for the observed period changes.  Averaged over a long enough time span, these noisy period changes would
presumably have to equal the rate of period change determined by evolution of the star through the HR diagram.  But
what is the origin of these noisy period changes?

Sweigart \& Renzini (1979) suggested that discrete mixing events in the semiconvective zone of an RR Lyrae star could
produce a period change noise with both increases and decreases of period.  In this model, the period changes
resulting from these discrete events would in the long run still equal those predicted by theoretical models that invoke
a smooth redistribution of chemical composition as the star evolves. 

Mass loss events, if sufficiently large, could
also produce period changes (Laskarides 1974).  However, calculations of models with plausible rates of mass loss do not
seem to produce period changes large enough to match observations (Koopmann et al. 1994). 

Cox (1998) proposed that
small changes in the gradient of the helium composition in the regions of RR Lyrae stars below the hydrogen and helium
convection zones might produce the noisy period changes. He suggested that an occasional dredge up of helium gravitationally settled
beneath the convection zones could be caused by convective overshooting. Cox applied this model to try to explain the
perplexing period changes in the double-mode RR Lyrae star V53 in M15 (Paparo et al. 1998), in which the fundamental mode
and the first overtone mode periods appear to change in different directions.  

Stothers (1980), considering the changes
in photospheric radius that might be associated with its magnetic energy content, suggested that hydromagnetic effects
might produce the period change noise. He later expanded on this idea in considering the explanation of the Blazhko effect 
(Stothers 2006).

Although some of these hypothetical causes of noisy period changes are more quantitative than others, it cannot
yet be proven which if any of the proposed mechanisms is correct.

\subsection{Period Changes in Globular Cluster RR Lyrae}

If it is difficult to test evolutionary theory from the period changes of an individual RR Lyrae star, perhaps we can
do better by looking at the patterns of period change observed in groups of RR Lyrae variables.  In this regard, the
RR Lyrae stars in globular clusters are of particular interest.  In the not-too-distant past, it was generally believed that all
of the RR Lyrae variables within a given globular cluster evolved from main sequence stars that formed at essentially
the same time and with the same initial chemical composition.  That may still be true for some globular clusters, but
others have recently been discovered to exhibit mutliple populations, so that the stars that they contain may not have been
as initially homogeneous as previously thought (Piotto 2009; Alonso-Garcia et al. 2013).  Nonetheless, there remain benefits from considering ensembles of RR Lyrae stars within globular clusters. The study of globular
cluster RR Lyrae stars is informed by our knowledge of the HR diagram and chemical abundances of the cluster in which they
occur.  Moreover, there is the non-negligible advantage that many cluster RR Lyrae stars can be observed on a single
photographic plate or CCD image (though, particularly in the days before the CCD, observations of the crowded RR Lyrae
near the cluster center could be difficult).

The observational record for RR Lyrae stars in several globular clusters now exceeds several decades, and for some
clusters extends to more than a century.  Often period changes are derived from various versions of the $O-C$
diagram. Large gaps in the observational records of some clusters complicate
the interpretation of $O-C$ diagrams, however.  As with field RR Lyrae stars, a gap in the observations can result in
a miscounting of the number of cycles that have elapsed between two observed epochs of maximum light, leading to
spurious values of period change.  That problem
is most serious for stars undergoing large and erratic changes in period.

Period change studies have been conducted for RR Lyrae stars in a number of the Milky Way's globular clusters, with some
clusters being studied several times as more data became available.  Though a complete listing of such studies is beyond
the scope of this review, we note recent studies for the Milky Way globular clusters M3 (Jurcsik et al. 2012), M5 (Szeidl et al.
2011), $\omega$ Cen (Jurcsik et al. 2001), M15 (Silbermann \& Smith 1995; Barlai \& Szeidl 1995), NGC 4147 (Stetson et al. 2005), 
NGC 5053 (Nemec 2004), NGC 7006 (Wehlau et al. 1992),  and IC4499 (Kunder et al. 2011).

Individual RR Lyrae stars within a single globular cluster exhibit the wide range of period change behavior seen among field RR Lyrae variables. Rathbun \& Smith (1997), found some evidence for a greater frequency of extreme
period changes among RRab variables than among RRcd stars.  This appeared to be true even when
the relatively small number of RRcd variables among most Oosterhoff I clusters was taken into account. As noted below, large period changes may be more common among RR Lyrae stars that show the Blazhko effect, a phenomenon also more common among RRab than RRcd stars.

The theoretical models of Lee (1991) predicted a correlation between the color distribution of HB stars within a
cluster and the average rate of period change $\beta$.  As shown in his Figure 4, the rate of evolutionary period change was predicted to
be small on average for RR Lyrae in most globular clusters, but would become increasing positive for globular clusters
with very blue distributions of HB stars.  In such clusters, RR Lyrae stars that begin their HB lives blueward of the
instability strip would evolve to the red through the instability strip on their eventual way to the asymptotic red giant
branch.  Results from those cluster period change studies included in Lee's Figure 4 appeared to confirm this general trend,
although the large increase in $\beta$ for clusters with blue horizontal branches depended mainly upon the single cluster
$\omega$ Centauri.  $\omega$ Cen is an unusual globular cluster, with a large star-to-star spread in heavy element
abundances.  It has been suggested that $\omega$ Cen might be the remnant of a dwarf galaxy absorbed into the
Milky Way (e.g. Lee et al. 1999).

An updated version of Lee's diagram (Figure 15 in Catelan 2009) is reproduced here as
Figure 4.  It employs new theoretical evolutionary calculations and adds clusters but shows a similar result. Once again $\omega$ Cen stands out with a value of $<\beta>$ significantly above those of most other clusters.  The average $\beta$ for NGC 4147 is also large, but is based on a smaller number of RR Lyrae observed over a shorter timespan than is the case for $\omega$ Cen, and
its value of $<\beta>$ consequently has a large uncertainty. To within the error bars, which are often considerable
compared to the small expected sizes of the evolutionary period changes, the mean values of $\beta$ appear to be broadly consistent
with the predictions of stellar evolution theory. However, the error bars are too large to quantitatively confirm
the exact values of the small $<\beta>$ values predicted for most globular clusters.

\begin{figure*}
\centering
\plotone{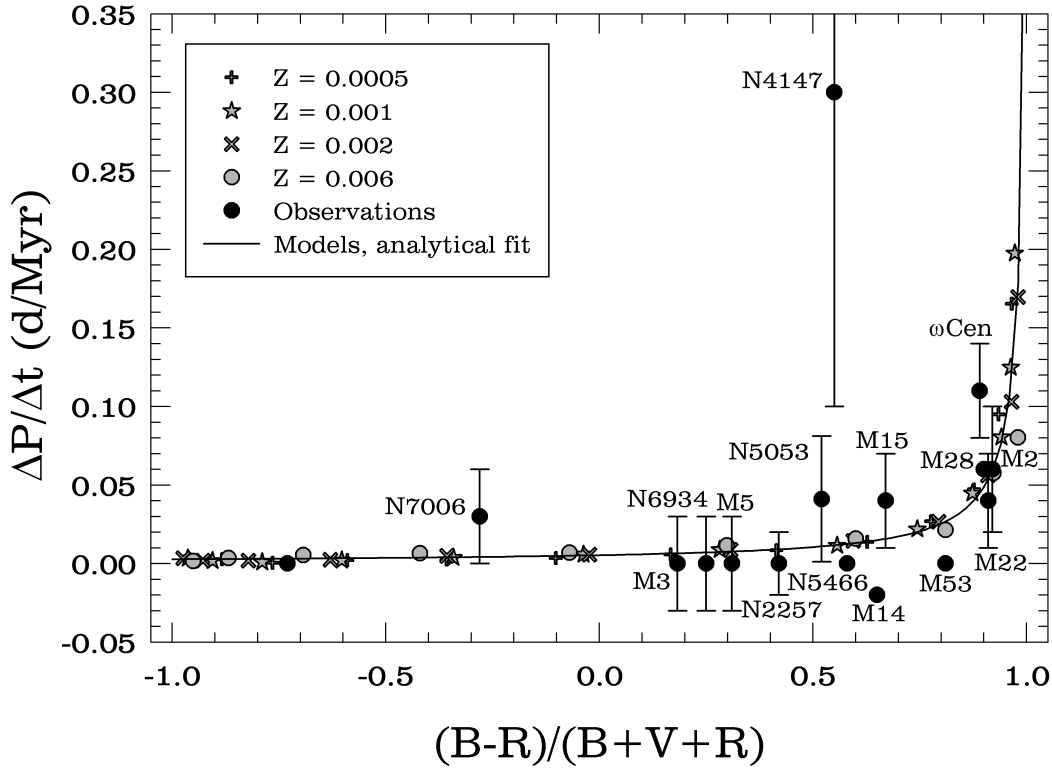}
\vskip0pt
\caption{This figure, from Catelan (2009), shows how the theoretically expected period changes of RR Lyrae stars (solid line) is a function of the distribution of stars across the horizontal branch.  B and R are the number of HB stars to the blue
and red side of the instability strip, respectively. V is the number of RR Lyrae variables.  Solid circles with error bars are
observed values for different globular clusters. }
\end{figure*}

The recent studies of RR Lyrae in the Oosterhoff I clusters M3 and M5 by Jurcsik et al. (2012) and Szeidl et al. (2011) deserve
particular attention.  These clusters have long and complete observational records, extending for about a century in the case
of M5 and 120 years in the case of M3, and
the treatment of the period changes in these papers is especially comprehensive. Period changes were determined
for 86 RR Lyrae stars in M5 and 134 RR Lyrae stars in M3. For M5 approximately two thirds
of the RR Lyrae star $O-C$ diagrams could be well fit by straight lines or parabolas (corresponding to constant or
linearly changing periods), but the remaining third had
$O-C$ diagrams indicating abrupt or variable period changes that were poorly represented by straight lines or quadratic terms.
In M3, the $O-C$ diagrams for 54 RR Lyrae stars were well fit by straight lines or parabolas, and those of an
additional 23 stars could be fit by straight lines or parabolas slightly less well.  However, the $O-C$ diagrams of 57 RR Lyrae
stars showed irregular variations. Thus, we again see that the observed period changes of many individual RR Lyrae
stars cannot be explained through stellar evolution alone, even if the average values of period change for all of the variables
within these clusters are consistent with the small value expected by evolution. Silva Aguirre et al. (2008) modeled the
behavior of pre-ZAHB stars in M3, finding that there should be about about 2 pre-ZAHB stars among the
M3 RR Lyrae variables.  They predicted that such stars would show large rates of period decrease, but too many such variables
are actually observed in M3.  

In both M3 and M5, Szeidl et al. (2011) and Jurcsik et al. (2012) found that irregular period changes were more likely among RR Lyrae
stars that exhibit the Blazhko effect, a long secondary periodicity in the light curve more common among RRab than RRc
stars.  The exact connection between the Blazhko phenomenon and period change remains unclear.  However, Szeidl et al. (2011) and Jurcsik et al. (2012) suggest that the observed period changes of Blazhko RR Lyrae may be less likely to represent evolutionary period changes than is the case for their non-Blazhko counterparts.  They conclude that, in averaging observed rates of period change in a cluster, one might calculate a value closer to the true evolutionary rate of period change by excluding those noisier RR Lyrae stars that have the Blazhko effect.  This result has important implications for the comparison of observed and theoretical rates of period change along the lines of Figure 4.

RR Lyrae stars of Oosterhoff type I and II can be distinguished by their positions in the Bailey (period-amplitude)
diagram. Sometimes, a few RR Lyrae stars in a mainly Oosterhoff I cluster will occupy the part of the Bailey
diagram normally filled by Oosterhoff II variables.  Such stars may be in a more advanced state of HB evolution than most of the RRab stars in the cluster. The same is true in reverse for RR Lyrae in the
Oosterhoff II cluster $\omega$ Cen. 

Kunder et al. (2011) examined whether the location of an RR Lyrae star within the Bailey
diagram was correlated with its period change. One might have expected the more evolved RR Lyrae stars to show larger rates of period change. However, they did not identify any difference between the period
change rates of RR Lyrae stars in the Oosterhoff I and Oosterhoff II regions of the Bailey diagram.  Jurcsik et
al. (2012) also detected no remarkable differences in period change rate for the M3 RR Lyrae stars whose
locations in the Bailey diagram suggested that they might be more highly evolved. Szeidl et al. (2011) noted that RR Lyrae stars in M5 did show some correlation between possible evolutionary state and observed period change.  Those M5 RR Lyrae that might be more evolved, based on their brighter apparent magnitudes and larger amplitudes, could exhibit large period increases but did not exhibit the large period decreases seen in possibly less evolved stars.

We are left with the result that the average period change rates for RR Lyrae stars within globular clusters
match the predictions of theory better than do the period changes of many individual RR Lyrae stars.  However, our ability
to test the theory of HB evolution with the existing period change results is limited.  The period change distributions
of RR Lyrae variables within individual globular clusters are poorly understood.      

\subsection{CCD Surveys}

Wide field CCD surveys that continue for long periods of time are opening up new avenues for the investigation of 
RR Lyrae period changes. As these studies expand in duration there is more opportunity for the
stars being monitored to undergo real period changes.  The southern part of the All Sky Automated Survey (ASAS, Pojmanski 2001)
has now been observing stars of $V$ = 8- 14 with full coverage south of +28 degrees since 2006 and with data
for a smaller area of the sky extending back further.  The various versions of the OGLE experiment 
(now up to OGLE IV) have monitored
the brightnesses of variable stars within parts of the Galactic bulge and Magellanic Clouds for two decades
(see http://ogle.astrouw.edu.pl/, and see also the paper by Welch et al. in this proceeding).  Wide field CCD
surveys are likely to become even more common in the future, with organizations such as the AAVSO becoming
involved in their operation.  Such surveys 
will undoubtedly play an increasingly important role in investigations of RR Lyrae period changes.

\section{Type II Cepheids}

Type II Cepheids are often divided into two subgroups based upon period.  Those with periods smaller than about 8 days are
often called BL Her stars while those with longer periods are named W Virginis stars (see, for example, Percy 2007). Some
use a different notation, terming above the horizontal branch variables with periods smaller than 3 days AHB1 stars (Sandage
et al. 1992).  Here, however, 
we shall consider all type II Cepheids with periods shorter than 8 days to be BL Her variables.  In the century that a
well-observed BL Her star may have been studied, it will have undergone some 10,000 or so cycles.  The corresponding number
for a typical W Vir variable is 2000 cycles.

The evolution of type II Cepheids is discussed by, among others,
 Gingold (1976), Bono et al. (1997) and Wallerstein (2002).  Type II 
Cepheids with mass $\le$ 0.59 solar masses are predicted to cross the instability strip
from blue to red on timescales of 1-2 million years (Bono et al. 1997).  However, type II Cepheids with masses as small as 0.52 or
0.53 solar masses may cross the instability strip 2 or 3 times as thermal pulses can send them on blue loops
into the instability strip.  The timescale for crossings in these cases is about 0.7 to 1.5 million years
(Bono et al. 1997).

If those theoretical calculations are correct, most BL Her stars would be evolving redward through the instability
strip toward
the asymptotic red giant branch (AGB).  By contrast, the longer period
W Vir type Cepheids may be low mass stars entering the instability strip during blueward loops.
We would thus expect BL Her variables to undergo period
increases while W Vir stars could show either period increases or period decreases, depending upon
the part of the thermal pulse loop in which they were located.  

The rate of evolutionary period increase expected for BL Her variables is
larger than that expected for RR Lyrae stars near the ZAHB and thus might be easier to observe.
BL Her variables are, however, far less numerous than RR Lyrae stars, so that the sample of stars available for period change studies is much
smaller.  Nonetheless, the results of period change studies for cluster (Osborn 1969;
Wehlau \& Bohlender 1982; Jurcsik et al. 2001) and field (Christianson 1983; Diethelm 1996; Provencal 1986) BL Her stars is more in line with the predictions of stellar evolution theory than is
true for RR Lyrae stars on an individual basis.  A majority of BL Her variables show period increases while the remainder show
no period change to within the observational uncertainty.  The rates of period change are broadly in line with theoretical rates.
As with the RR Lyrae variables, the period changes observed over a span of a century are small compared to the periods themselves, but 
the observed rates of period change can be as large as about 20 days per million years.  The $O-C$ diagrams for these stars
are often inadequate to tell whether the stars change period at a constant rate or whether the period changes happen in
abrupt steps.

Unlike the BL Her variables, W Vir stars in both clusters and the field can show significant period decreases as well as
increases or they can be observed to be constant in period to within the uncertainties
 (Percy \& Hale 1998; Percy et al. 2000; Templeton \& Henden 2007; Rabidoux et al. 2010; Coutts Clement \& Sawyer Hogg 1977; Berdnikov
\& Pastukhova 2007; Clement et al. 1988).  The magnitude of the observed period changes is broadly consistent with
evolutionary theory.  However, when we consider the longest period W Vir stars we find that their light curves can become increasingly
unstable, taking on aspects of the more irregular variations seem among RV Tauri stars (Rabidoux et al. 2010; Percy \&
Coffey 2005). This is not true of all long period W Vir stars, as can be seen in a comparison of V42 and V84 in the globular cluster M5, which have
periods near 25 days.  Despite their similar periods, the $O-C$ diagram of V42 shows much less erratic variation than that of V84 (Rabidoux et al. 2010).

Soszynski et al. (2008; 2010; 2011) used OGLE data to investigate Cepheids within the Magellanic Clouds and the Galactic bulge.  Data for SMC variables were obtained over an 8 or 13 year interval.  Data for LMC variables were obtained over 7 years, while the bulge data were obtained between 1997 and 2009. These surveys now extend over an interval long enough that significant period changes have been detected among some of their type II Cepheids.  Soszynski et al. (2011) noted that the rate of period change in the W Vir star OGLE-BLG-T2CEP-059 was so rapid that its light curve could not be phased using a constant period.  As the duration of the OGLE surveys increases, they will become increasingly important in the study of type II Cepheid period changes.

\section{Summary}

The largest period changes observed in a small number of Mira variables are consistent with those expected from thermal pulses, although it
is not certain that such pulses are responsible for all large period changes in Mira stars.  The smaller period
changes observed among a larger number of Mira variables do not have a fully satisfactory explanation.
Period changes of type II Cepheids appear consistent with the predictions of stellar
evolution theory.  By contrast, the period changes of individual RR Lyrae stars are often larger
and more changeable in sign and magnitude than stellar evolution theory would predict.  The average period changes for
all RR Lyrae stars within a globular cluster provide a closer match to evolutionary expectations. Excluding Blazhko variables from the average might make it easier to discern the evolutionary rate of period change. Period changes that amount to a significant fraction of the period of the star are very rare and are restricted to a few Mira variables.  Period changes observed among RR Lyrae stars or type II Cepheids do not amount to more than a few parts in $10^4$ of the periods themselves.

\section{Acknowledgements}

I would like to thank Marv Baldwin (field stars) and the late Adriaan Wesselink (cluster RR Lyrae)
for introducing me to the study of RR Lyrae period changes. I also thank my former students for all the work they put into organizing this conference. I thank Matthew Templeton and Marcio Catelan for their comments and for supplying updated figures for this paper.  I am grateful to Johanna Jurcsik for valuable comments on an earlier draft of this paper.


\end{document}